\documentstyle[aps,prd,multicol,epsfig]{revtex}
\input{psfig.tex}
\begin{document}

\newcommand{\tl}{\tilde}
\newcommand{\bm}{\boldmath}
\newcommand{\cut}{\rm cut}
\newcommand{\mnras}{Mon. Not. Roy. Astron. Soc.}
\def\bi#1{\hbox{\boldmath{$#1$}}}
\def\sun{\hbox{$\odot$}}


\title{Sunyaev-Zeldovich effect from hydrodynamical simulations: 
maps and low order statistics}

\author{Uro\v s Seljak,
Juan Burwell}
\address{
Department of Physics, Jadwin Hall, Princeton University,
	Princeton, NJ 08544
}

\author{Ue-Li Pen}
\address{Canadian Institute of Theoretical Astrophysics, University of
Toronto, 60 St. George St., Toronto, Canada; pen@cita.utoronto.ca}

\date{Submitted to PRD - January 2000}

\maketitle
\begin{abstract}
We use moving mesh hydrodynamical simulations to make maps of Sunyaev-Zeldovich 
effect. We present these maps for several cosmological models and explore
their lowest moments. We find that the first moment, the mean Compton $y$ 
parameter, is typically between $1-2\times 10^{-6}$ for cluster abundance 
normalized models, the lower value corresponding to the high density models, 
and scales as $\sigma_8^{3-4}$. Rms fluctuations at 10' scale have an 
amplitude $ \Delta T/T \sim 1-3 \times 10^{-6}$ in the Rayleigh-Jeans regime. 
The amplitude of 
the power spectrum strongly depends on the power spectrum normalization and 
scales as $\sigma_8^{6-9}$. On smaller scales ($l>1000$) the spectrum is 
dominated by halos below $10^{13}M_{\sun}$ and so is sensitive to the thermal 
history of galactic halo gas. On larger scales ($l<1000$) the power spectrum 
is less sensitive to nongravitational energy injection, but becomes very noisy,
with large field to field variations in the power spectrum caused by rare 
bright sources in the maps, which dominate in the spectrum over the large scale 
structure correlations. Cross-correlation power spectrum with weak lensing or 
projected galaxy distribution is significant and the cross-correlation 
coefficient is around 0.5 over a wide range of scales. Comparison with 
Press-Schechter predictions gives very good agreement for all the statistics, 
indicating that there is no significant contribution to SZ from non-virialized 
structures. The one point distribution function shows clear deviations from 
gaussianity and can be well-approximated as log-normal on small scales.
\end{abstract}

\section{Introduction}
\label{intro}

Cosmic microwave background (CMB) photons propagating through the universe are 
scattered by hot electrons along their path. The effect of this 
scattering on the photon distribution was first described by 
Sunyaev and Zeldovich \cite{sz70} and is called the Sunyaev-Zeldovich (SZ)
effect. The effect conserves the number of photons, but changes their 
energy distribution. Low frequency photons in the Rayleigh-Jeans 
regime gain energy on average and are moved into the high frequency
part of the
Planck distribution, with a zero crossing at 217GHz in the nonrelativistic
case. The amplitude of the effect is proportional to the 
product of electron temperature and density, or pressure, and is thus 
dominated by hot dense structures such as clusters. 

The SZ effect is nowadays routinely detected by a number of instruments
such as BIMA, Diabolo, OVRO, PRONAOS, Ryle, SEST, SuZie and Viper (see
refs. \cite{rep95,bir99} for review). 
All of these have so far concentrated their efforts on known clusters, 
but future experiments will also observe blank fields of the sky with 
an effort to isolate this effect. Among these are the proposed one 
square degree survey \cite{Carlstrom99} and a number of CMB 
experiments, for which SZ may be a significant source of fluctuations that 
needs to be separated from the primary signal. The Planck 
satellite for example, with its combination of frequency coverage, 
all-sky map and angular resolution, should be able to identify several
thousand SZ sources \cite{Aghanim}. 

The SZ effect has a number of potentially observable consequences in 
addition to the one descibed above on the known clusters. First, the 
mean SZ distortion 
can be observed through the deviation of the 
photon spectrum from the Planck distribution. No such distortion has 
been detected by the COBE/FIRAS data yielding the upper limit on 
Comptonization parameter $y$ (defined below) of $y<1.5 \times 10^{-5}$
\cite{Fixsen96}. This is a constraint that has to be satisfied by 
any viable cosmological model. As we will show below cluster abundance
normalized models explored here typically satisfy this constraint. 

Second, most of the CMB experiments measure fluctuations, 
for which the most relevant quantity is their power spectrum.
Again no clear detection of SZ has been reported so far, 
although this is not unexpected given that the main source of 
fluctuations on degree scales probed so far are primary CMB
fluctuations generated at the time of recombination. 
The next generation of CMB experiments is beginning to 
detect fluctuations 
at smaller angular scales. With higher sensitivities and 
broader frequency coverage SZ may 
become detectable even in random patches of the sky surveyed by these
experiments. There are two related questions that one would like to 
address. 
First is the overall level of SZ fluctuations and whether
SZ may be a significant source of CMB foreground contamination. Second 
is the power of SZ to
distinguish between cosmological models using power 
spectrum or any other statistics, such as counting of sources or nongaussian
signatures. 

There are two main theoretical approaches that can be used when calculating the 
SZ effect. One is analytical and is based on the Press-Schechter 
approximation. In this approach all the mass in the universe is 
distributed into virialized halos. Gas in these halos is assumed to 
follow a prescribed distribution in density and temperature (usually 
isothermal with a $\beta$ density profile). One then integrates over
the mass function and over the redshifts 
to calculate the first moment of the distribution (the $y$ parameter).
A similar calculation also yields the uncorrelated contribution 
to the second moment \cite{Atrio99}
to which one must also add the correlations 
between the halos \cite{ColeKaiser88,KomatsuKitayama99}. 
This approach involves 
a number of approximations that must be verified with 
numerical simulations. For example, some fraction of the gas may not be 
in virialized halos, but may be residing in more diffuse filaments or 
at the outskirts of the halos. There may be significant temperature and 
density fluctuations of the gas inside the halos. Furthermore,  
N-body simulations that have examined the accuracy of the
Press-Schechter mass function have shown it to be a reasonable 
approximation for massive halos, but to overpredict the number of 
halos at lower masses. The resulting spectrum may be very sensitive to the 
details of the mass function distribution.

The alternative approach adopted in this paper is to use numerical 
simulations to calculate the SZ effect. Our approach is 
based on moving mesh hydrodynamical simulation code \cite{Pen97} and we use 
ray tracing through the simulation box to generate maps of SZ.
This is the approach that 
was adopted previously by \cite{ThomasCarlberg89,Scaramella93}. 
Most of these early 
studies were of low resolution, but nevertheless showed that SZ could
be an important effect on the CMB. Recently \cite{Silva99}
generated SZ maps using SPH simulations of higher resolution, 
although they did not calculate the CMB power spectra. 
An alternative method based on $3-d$ power spectra was developed in
\cite{persi95} and was recently adopted to the 
MMH code as used in this paper by
\cite{refregier99}.
We compare our results to the previous work where possible. 

\section{Method}
We use moving mesh hydro (MMH) code developed by one of us
\cite{Pen97}. The code 
is a hybrid between Eulerian and Lagrangian grid based hydrodynamic 
methods. By deforming the grid in the dense regions along potential flow lines
it provides a ten fold increase in resolution compared to fixed 
grid Eulerian codes, while maintaining regular grid conditions everywhere
\cite{Pen97}. The code has been succesfully parallelized on shared memory
systems and has a low computational cost per grid cell. A $256^3$ run 
takes around 2-4 days of wall clock time on 32 processors. 

We ran 
about a dozen simulations varying cosmological parameters, box sizes and 
particle/mesh dimensions to check for various numerical effects. 
Our main models are $\tau CDM$, $\Lambda CDM$ and $O CDM$.
Their parameters are
\begin{itemize}
\item $\tau CDM$: flat model with 
$\Omega_m=1$,  
$\Omega_{\Lambda}=0$, $\sigma_8=0.56$, 
$\Omega_b h=0.034$ and $\Omega_m h=0.25$. 
\item $\Lambda CDM$: flat model with 
$\Omega_m=0.37$, $\Omega_{\Lambda}=0.63$, $\sigma_8=0.8$, 
$\Omega_b h=0.034$ and $\Omega_m h=0.25$. 
\item $O CDM$: open model with 
$\Omega_m=0.37$, $\Omega_{\Lambda}=0.0$, $\sigma_8=0.8$, 
$\Omega_b h=0.034$ and $\Omega_m h=0.25$. 
\end{itemize}

We ran several simulations varying box sizes from 50$h^{-1}$Mpc to 
200$h^{-1}$Mpc to verify the effects of mass and force 
resolution on small scales and lack of power on large 
scales. We also varied mass and scale resolution, using mesh 
sizes between $128^3$ and $256^3$ and typically placing 1-8
dark matter particles per mesh cell. Our largest simulations are 
$\Lambda CDM$ and $OCDM$ models with 100$h^{-1}$Mpc box, $256^3$ mesh and 
$256^3$ dark matter particles.

During the simulation we store 2-d projections through the 3-d box
at every conformal 
time step that corresponds to a light crossing time through the 
box. The projections are made alternatively in $x$, $y$ and $z$ 
direction to minimize the repetition of the same structures in 
projection. We store projections of SZ, kinetic SZ, gas and 
dark matter density. For SZ we store the projection 
\begin{equation}
\Delta y={k_B\sigma_T \over m_ec^2}n_eT_e a\Delta \chi 
\label{dy}
\end{equation}
where 
$\sigma_T$ is the Thomson cross-section, c the speed of light, 
$m_e$ the electron mass, $\Delta \chi$ the comoving width of the box, 
$a$ the expansion factor, 
$n_e$ electron number density and $T_e$
electron temperature, both of which are obtained from the output of 
the MMH hydro code (note that we ignore relativistic corrections, which 
are only relevant for rare hot clusters). 
Our 2-d maps are $1024^2$ for $256^3$
and $512^2$ for $128^3$. We have verified that this preserves all 
the information content by comparing the results 
to the higher resolution projection. 
Number of stacked projections depends on the box size and ranges between 
25 (for $\tau CDM$ with 200$h^{-1}$Mpc box) to 150 (for $\Lambda CDM$ with 
50$h^{-1}$Mpc box size). 

After the simulation is completed we use the 2-d projections to 
make maps of SZ. We stack together the maps of SZ separated by 
the width of simulation box, randomly choosing the center of each box 
(note that use of periodic boundary conditions guarantees there is no 
edge for any of the maps). We then project these maps onto a map
constant in angular size. In principle the angular 
scale of the projection should be determined by the angular scale of the 
simulation 
box at the initial $z$, since any larger scale produces repetition of 
the same structures in the map.
However, most of the structures in the maps are
coming from low $z$ and these cover very little volume of the simulation box; 
hence they do not repeat itself even if we increase the angular scale beyond
the size of the box at the highest redshift.
Typically we project up to $z \sim 4-30$ 
using oversampling between 1-4 without any noticeable artifacts in the maps.
We checked for possible artifacts in the power spectra by comparing 
weak lensing maps produced in the same way 
to the analytic predictions \cite{JainSeljak98,JSW99}. 
The agreement was 
very good in all models. For SZ it should be even better, since SZ power 
spectrum is much more dominated by isolated sources and, as will be 
shown below, large scale correlations do not dominate even on large scales.

A given simulation can be used not only for the simulated parameters, but 
also for models with lower $\sigma_8$. To do this one can simply relabel an 
earlier time output as being one at a later time by changing 
the redshift of projection 
\begin{equation}
z^{\rm new}=(z^{\rm old}+1)\sigma_8^{\rm new}/
\sigma_8^{\rm old}-1,
\label{res}
\end{equation}
where superscript old stands for original simulation 
and new for the new one. Once the redshifts are relabeled one can 
project over the 2-d projections using the nearest projection to the
required $z$. For a flat model with no cosmological constant 
this scaling preserves all cosmological parameters (except $\sigma_8$). 
This scaling is particularly useful to obtain the 
dependence on $\sigma_8$ in a flat model independent of other parameters. 
We verified its accuracy by 
running two simulations: one with $\sigma_8=1$ but using the rescaling of 
equation \ref{res} to obtain $\sigma_8=0.56$ model, and another  
with $\sigma_8=0.56$ at the final output, and found very good
agreement between the two. 
The simulation with $\sigma_8=1$ was then used to predict 
$\sigma_8$ dependence of the $y$ parameter and SZ power spectrum. 

It is also easy to verify the effect of the baryon density in the 
limit $\Omega_b \ll \Omega_m$. In this limit the
baryons are just a tracer
of dark matter potential and do not dynamically couple to it.
This fails in the cores of the halos, which are however already not 
correctly modelled in these adiabatic simulations without
cooling and non-thermal energy injection. In this approximation 
SZ scales linearly with baryon density times Hubble constant $\Omega_b h$.
We only show results that are normalized to BBN nuclesynthesis constraint 
$\Omega_bh^2=0.018$, but results for other values of $\Omega_b h$ can 
be easily obtained using this scaling.
Incidentally, one could also use  
transformation of scale to map 
a given model into another family of models, this time varying the 
shape and the amplitude 
of the input power spectrum. By combining this transformation with the 
time transformation above one can find a family of models which are 
normalized to $\sigma_8$ today and have a different shape of the power 
spectrum. 
The trade-off in this case is a loss in force and mass resolution. 
Some of the maps produced by the simulations can be found at 
http://feynman.princeton.edu/$\sim$uros/sz.html. 

\section{Press-Schechter predictions}

The Compton parameter $y=\int dy$ 
is given as a projection over the electron pressure along 
the line of sight (equation \ref{dy}). This can be reexpressed as 
a line of sight integral over the density weighted temperature 
of electrons, 
\begin{equation}
y={k_B\sigma_T \bar{n_e}\over m_ec^2}\int a^{-2}(1+\delta_g)T_e d\chi,
\label{yd}
\end{equation}
where $\delta_g$ is the gas overdensity and $ \bar{n_e}$ mean electron 
density today. 

In Press-Schechter (PS) picture all matter in the universe is divided 
into halos of a given mass. The mass distribution is specified 
by the halo density mass function $dn(M)/dM$.  This can be 
written as 
\begin{equation}
{dn(M) \over dM} dM={\bar{\rho} \over M}f(M)dM, 
\end{equation}
where $\bar{\rho}$ is the mean matter density of the universe. The function 
$f(M)$ denotes the fraction of mass in halos of mass $M$. It can be 
expressed in units in which it has a universal form independent of 
power spectrum or redshift if we express
it as a function of peak height $\nu=[\delta_c(z)/\sigma(M)]^2$, where
$\delta_c$ is the value of a spherical overdensity at which it 
collapses at $z$ ($\delta_c=1.68$ for Einstein-de Sitter model) and 
$\sigma(M)$ is the rms fluctuation in spheres that contain on average
mass $M$ at initial time, extrapolated using linear theory to $z$ .
For scale free spectra
\begin{equation}
\nu f(\nu)= {M^2 \over \bar{\rho}}{dn \over dM} { d\ln M \over d\ln
\nu}. 
\label{eqn:ps}
\end{equation}
The actual form given by Press \& Schechter \cite{ps74} is  $\nu f(\nu)=
(\nu/2\pi)^{1/2}e^{-\nu/2}$, although modified versions of this form 
that fit better N-body simulations have been proposed
\cite{ShethTormen99}.  Direct hydrodynamic simulations have shown good
agreement with our form \ref{eqn:ps} \cite{pen98}.

The virial temperature of the halo in the spherical collapse model 
is only a function of virial mass and is given by $k_BT_e = q M^{2/3}$. 
The conversion factor from mass to temperature $q$ is approximately 
unity if mass is expressed in units of $10^{13}h^{-1}M_{\sun}$ and 
$T$ in keV.
Density weighted temperature is given by 
\begin{equation}
\langle (1+\delta)T\rangle = q \int f(\nu)d\nu M^{2/3}.
\label{rhot}
\end{equation}
This gives $\langle (1+\delta)T\rangle =0.3$keV today in a $\Lambda CDM$ 
model with $\Omega_m=0.37$ and $\sigma_8=0.8$, with other models
giving comparable numbers. Further assuming that gas traces dark matter, 
ie $\delta_g=\delta$, we can calculate the integrated Compton $y$ parameter
from equation \ref{yd}. This gives $y=2\times 10^{-6}$ for 
this model. Flat models give a factor of 2 lower value because of
a more rapid evolution of clusters with $z$. This 
reduces $y$ relative to a
low density model assuming the cluster abundance today 
is the same. 

Using equation \ref{rhot} we may also explore the dependence of $y$
on $\sigma_8$. For this we need to relate $\sigma(M)$ to
$M$. Linear power spectrum at the cluster scale can be 
approximated as a power law $P(k) \propto \sigma_8^2 k^n$.
Since rms variance $\sigma^2(M)$ scales as $ k^3P(k)$ one finds 
$k \propto \nu (\sigma_8^2)^{-2/(n+3)}$. Mass goes as 
$M \propto \bar{\rho}R^3 \propto k^{-3}$ for which
we find $M^{2/3} \propto k^{-2} \propto \sigma_8^{{4 \over 3+n}}$. 
On cluster scales we have $-2<n<-1$, hence 
$\langle (1+\delta)T\rangle \propto \sigma_8^{2-4}$. 
The $y$ dependence on $\sigma_8$ depends on the projection of 
$\langle (1+\delta)T\rangle$ along the line of sight 
(equation \ref{yd}), but 
gives qualitatively similar result. 

One can also compute power spectrum using the PS model. To do this one
can first calculate the 3-d power spectrum of density weighted temperature 
$P_{(1+\delta) T}(k)$ 
and then project it along the line of sight using the Limber's equation, 
\begin{equation}
C_l=32\pi^3\left[{k_B\sigma_T \bar{n_e}\over m_ec^2}\right]^2\int 
P_{(1+\delta) T}(k=l/r,a) {d\chi \over a^4 r^2},
\label{limber}
\end{equation}
where $r$ is the comoving angular distance to $\chi$, given by
$\chi$, $R\sinh(\chi/R)$ or $R\sin(\chi/R)$ in a flat, open or closed
universe with curvature R, respectively. The above expression
is valid in Rayleigh-Jeans limit where $\Delta T /T \equiv j(x)y=-2y$. 
This can easily 
be rescaled to another frequency using the spectral function 
$j(x)=x(e^x+1)(e^x-1)^{-1}-4$ with $x=h\nu/k_BT$ (which in the limit $x
\rightarrow 0$ gives $j=-2$).
 
In the PS formalism there are two contributions to the pressure power spectrum 
\cite{Atrio99,ColeKaiser88,KomatsuKitayama99}
\begin{equation}
P(k)=P^P(k)+P^{hh}(k)
\end{equation}
The first term $P^P(k)$ arises from the correlations
within the single halo.
This term contribution to the 3-d power spectrum is given by 
\begin{equation}
P^P_{(1+\delta)T}(k)= {q^2 \over (2\pi)^3}\int f(\nu)d\nu {M \over \bar{\rho}}
M^{4/3}|y(k)|^2,
\end{equation}
where $y(k)$ is the Fourier transform of the halo profile normalized to 
unity on large scales ($k \rightarrow 0$), assumed here for simplicity to 
be independent of $M$.
One power of $M$ above is given by mass pair weighting and $M^{4/3}$ is given 
by the square of the temperature of the halo. This term is heavily 
weighted toward rare massive clusters, resulting in a very strong 
Poisson term compared to the halo-halo correlation term discussed below. 
As shown in \cite{KomatsuKitayama99}
this term dominates over the halo-halo correlation 
even on large scales, where it behaves as a white noise. In the 
large scale limit ($k \rightarrow 0$, $y(k)=1$) 
the scaling with $\sigma_8$ is given by
$P^P_{(1+\delta)T} \propto \sigma_8^{14/(3+n)} \propto \sigma_8^{7-14}$. This 
will again be modified somewhat by the integration over the redshift, but
does provide qualitative understaing for the strong 
$\sigma_8$ dependence seen in the
simulations described below. 

Second term is the contribution from halos correlated 
with one another. On large scales these cluster according to the linear 
power spectrum $P_{\rm lin}(k)$, 
except that they can be biased relative to the dark matter. 
The halo bias can be either 
larger than unity for halos more massive than the nonlinear mass or 
less than unity for those below that. An expression that fits N-body 
simulations reasonably well was given in \cite{ColeKaiser88,MoWhite96} 
(see also \cite{ShethTormen99,Jing98} for a 
modification relevant for lower mass halos)
\begin{equation}
b(\nu)=1+{\nu -1 \over \delta_c}.
\end{equation} 
The halo-halo contribution to the density weighted $T$ power spectrum 
is given by 
\begin{eqnarray}
P^{hh}_{(1+\delta)T}(k)&=&P_{\rm lin}(k)\left[q\int 
f(\nu)d\nu M^{2/3}(\nu) b(\nu)y(k)\right]^2  \nonumber \\
&\equiv & P_{\rm lin}(k)\left[\langle b\rangle_{(1+\delta) T}\langle(1+ \delta) T \rangle 
\right]^2, 
\end{eqnarray}
where $\langle b\rangle_{(1+\delta) T}$ is the pressure weighted bias, defined 
as  
\begin{equation}
\langle b\rangle_{(1+\delta) T} \equiv {\int f(\nu)d\nu M^{2/3}(\nu) b(\nu) y(k)\over 
\int f(\nu)d\nu M^{2/3}(\nu)y(k)}.
\label{bias}
\end{equation}
For $\Lambda CDM$ model we find in the large scale limit ($y=1$)
$\langle b\rangle_{(1+\delta) T}=3$,
hence SZ halos are significantly biased relative to the dark matter.
This is because the $T \propto M^{2/3}$ weighting weights preferentially 
towards the large halo masses which are biased. 
The bias decreases towards smaller scales where $y(k)$ suppression for finite 
$k$ is more important for larger more massive halos and 
the dominant contribution shifts towards smaller, unbiased or 
antibiased halos. Despite this large bias the halo correlation term is 
small relative to the halo term even on large scales and Press-Schechter 
model predicts that SZ power spectrum does not trace large scale structure
on large scales.

If SZ does not trace LSS, the large scale bias $b$ cannot be measured from 
its power spectrum. Cross-correlating SZ
with weak lensing or galaxy map is a more promising 
way to obtain this large scale bias. This is because the Poisson term is 
given by $M^{5/3}$ weighting,
\begin{equation}
P^P_{(1+\delta)T,\delta}(k)= 
{q \over (2\pi)^3}\int f(\nu)d\nu {M \over \bar{\rho}}
M^{2/3}|y(k)|^2,
\end{equation}
and is so less dominated by rare objects 
than SZ power spectrum with $M^{7/3}$ weighting. 
The correlated contribution to the cross-correlation
between density weighted $T$ and 
dark matter density is given by 
\begin{eqnarray}
P^{hh}_{(1+\delta)T,\delta}(k)&=&P_{\rm lin}(k)\left[q\int f(\nu)d\nu M^{2/3} b(M)\right]
\left[\int f(\nu)d\nu  b(M)\right]  \nonumber \\
&=&P_{\rm lin}(k)\langle b\rangle_{(1+\delta) T}\langle (1+\delta) T \rangle 
,
\end{eqnarray}
where the second integral above is the mass weighted bias,
which is by definition unity. Numerical comparison between the two terms 
indeed shows that the are comparable on large scales.
To obtain the SZ-weak lensing or galaxy cross-correlation 
power spectrum we again use this 3-d power spectrum in Limber's equation 
\ref{limber} with the appropriate window function for the dark matter 
or galaxy projection. 

\section{Simulation results}

\subsection{Mean Compton parameter}
We first present the results on the mean $y$ parameter. The value 
for this parameter is 
$1.0 \times 10^{-6}$ for $\tau CDM$
in 100$h^{-1}$Mpc simulations with $128^3$ mesh, 
$2.4 \times 10^{-6}$ for 
$O CDM$ in 100$h^{-1}$Mpc with $256^3$ mesh and
$2.0 \times 10^{-6}$ for $\Lambda CDM$ with 100$h^{-1}$Mpc with $256^3$ mesh or
50$h^{-1}$Mpc with $128^3$ mesh. 
For the latter $y$ decreases by roughly 20\% if  
the resolution is decreased to $128^3$ mesh, 
suggesting that small halos not resolved in larger box simulations 
contribute a significant fraction to its 
value. This is also confirmed by Press-Schechter type calculations 
\cite{Atrio99}. Non-thermal energy injection 
and cooling may provide additional corrections to the results of 
the adiabatic simulations. For example, \cite{CO} include
feedback in their simulations and find 
significantly higher density averaged temperature (1keV) than 
our simulations (0.3keV), although other simulations that also 
include heating do not (G. Bryan, private communication). 
The mean $y$ parameter is therefore sensitive to the thermal 
history of the gas.

Our results are still 
one order of magnitude below the current experimental limits from 
COBE/FIRAS $y < 1.5 \times 10^{-5}$ \cite{Fixsen96}, so it is unlikely 
that this constraint will play a major role in distinguishing between 
the models that differ in the history 
of energy injection in the universe.
On the other hand mean $y$ parameter does put a constraint on the $\sigma_8$. 
This scaling of $y$ with $\sigma_8$ can be obtained from the same simulation 
using the method described in previous 
section. We find $y \propto \sigma_8^{3-4}$, which is in a good agreement with 
the Press-Schechter predictions. To violate the FIRAS limits one needs to 
increase $\sigma_8$ by a factor of 2 over the cluster abundance value.
COBE normalized standard CDM with $\sigma_8=1.3$ would be problematic.
Our results are a factor of 2 lower from those in \cite{Silva99} 
for the two low
density models (once we 
account for the difference in $\Omega_b h$), while we are in a good
agreement for $\tau CDM$ model. 
This disagreement does not seem
to be explained by 
limited mass resolution on small scales, 
since our simulations have higher mass and force
resolution, yet predict lower values for $y$ for the low density models. 
We agree well with results in \cite{refregier99}.

\subsection{SZ power spectrum}
\begin{figure}
\centerline{\psfig{file=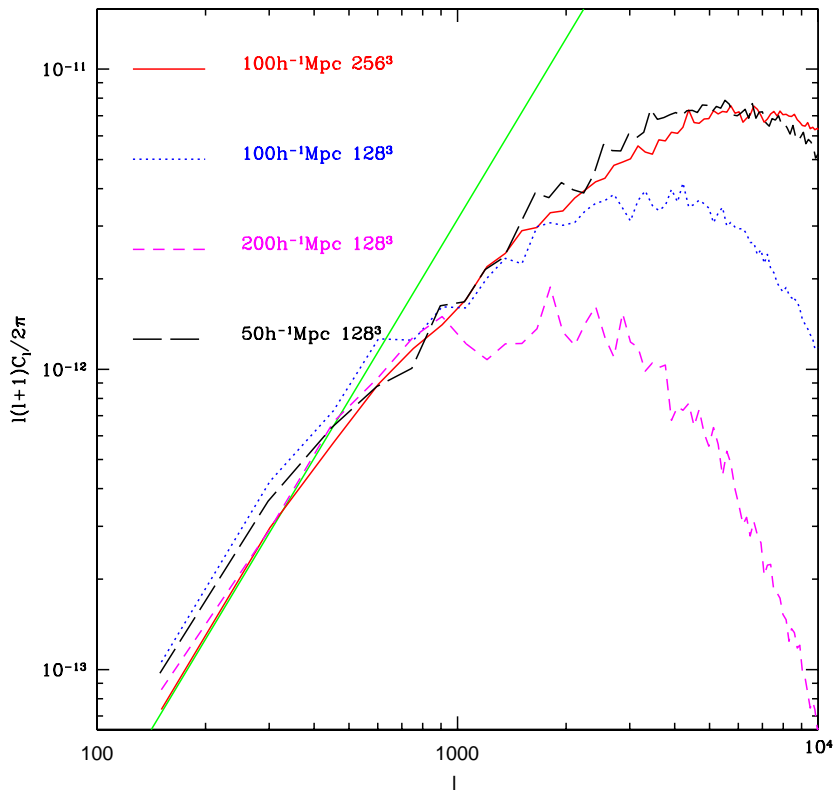,width=4.5in}}
\caption{
Power spectrum 
comparison between different resolution and box size simulations for 
$\Lambda CDM$ model. Also shown is the white noise spectrum scaling 
as $l^2$. On large scales the SZ spectrum is white noise in all cases.
}
\label{fig1}
\end{figure}

Next we explore the power spectra of SZ in the Rayleigh-Jeans regime. 
We first investigate the effects of resolution.
These are shown in figure \ref{fig1} for different parameters of the 
simulation for $\Lambda CDM$ model. Shown are $256^3$ 100$h^{-1}$Mpc 
simulation, $128^3$ 100$h^{-1}$Mpc simulation, $128^3$ 200$h^{-1}$Mpc simulation
and $128^3$ 50$h^{-1}$Mpc simulation. They all agree on large scales,
where the shape is close to the white noise model, characteristic of 
a power spectrum dominated by rare sources.
On small scales
200$h^{-1}$ simulation begins to loose power
for $l > 1000$ compared to other simulations. 
Comparison between $128^3$ and $256^3$ 
100$h^{-1}$Mpc shows that the two are in good agreement on scales up to 
$l \sim 1500$, beyond which $128^3$ begins to loose power. Similar 
conclusion is obtained by comparing 50$h^{-1}$Mpc and 100$h^{-1}$Mpc
both with $128^3$. The figure shows that a 50$h^{-1}$Mpc box is sufficient 
for the power spectrum on large scales. This indicates that the large 
scale power spectrum is not dominated by very massive clusters above 
$10^{15}M_{\sun}$ which do not form in such small boxes, but rather 
by less massive 
$10^{14-15}M_{\sun}$ clusters, which happen to be nearby in the projection
along the line of sight. This conclusion is also confirmed by the
Press-Schechter calculation, where the brightest sources correspond 
to this mass range. 
On small scales the 50$h^{-1}$Mpc box has a power spectrum comparable to 
$256^3$ 100$h^{-1}$Mpc 
simulation with the same mass and force resolution. 
Very little small 
scale power therefore arises from mode-mode coupling 
with scales larger than $50h^{-1}$Mpc. 
Overall 100$h^{-1}$Mpc
$256^3$ simulation resolves the power spectrum between $100<l<5000$,  
while $128^3$ simulations of the same size are sufficient between 
$100<l<2000$. 

\begin{figure}
\centerline{\psfig{file=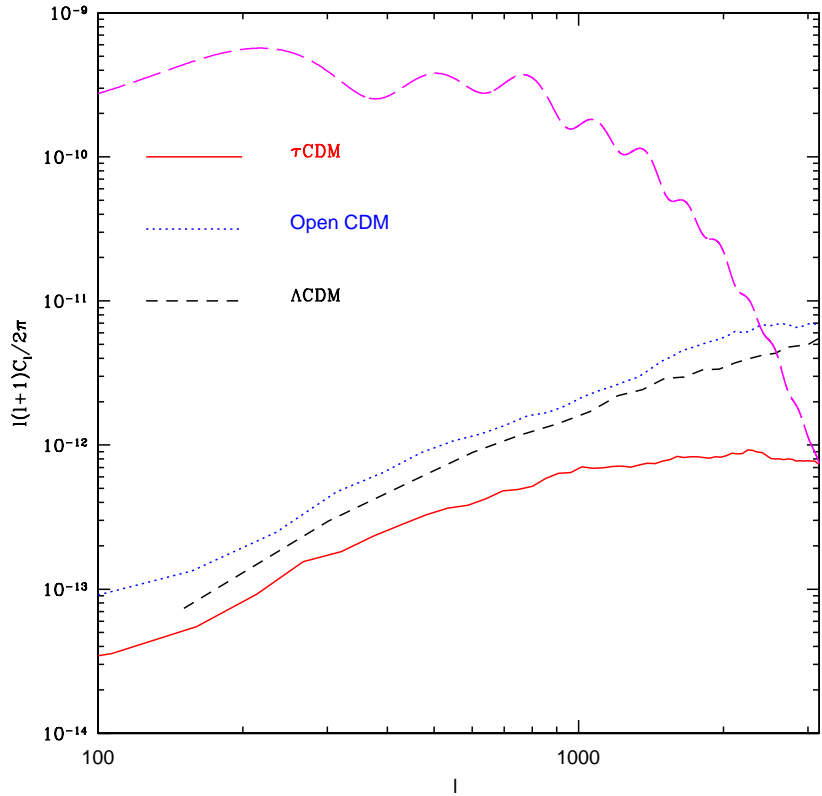,width=4.5in}}
\caption{
Power spectrum comparison between  
cluster normalized $\tau CDM$,
$\lambda CDM$ and $OCDM$. The latter two have significantly more 
power. Also shown is the primary CMB (long dashed) which dominates
on large scales. 
}
\label{fig2}
\end{figure}

On large scales the power spectrum approaches white noise in slope. 
This is an indication that the power spectrum is dominated by the 
rare bright sources in the map and not by the correlations between 
them, which would give a much shallower slope. The same behaviour
is seen also in other cosmological models, shown in figure \ref{fig2}.
While $OCDM$ gives similar 
predictions to $\Lambda CDM$ model, $\tau CDM$ is significantly lower, 
just like in the case of mean $y$ parameter. Low density $\Lambda CDM$
predicts
$3 \times 10^{-7}$ fluctuations around $l \sim 100$, 
which rises to 3$ \times 10^{-6}$
at $l \sim 5000$. Comparison with primary CMB in figure \ref{fig2}
shows that the SZ is unlikely to contaminate CMB  
power spectrum for MAP, while Planck and smaller scale 
experiments should be able to measure the SZ power spectrum.
Our results are in good agreement with 
\cite{refregier99}, 
although we note that our $OCDM$ spectrum is somewhat higher than theirs,
in better agreement with the PS calculations. 

\begin{figure}
\centerline{\psfig{file=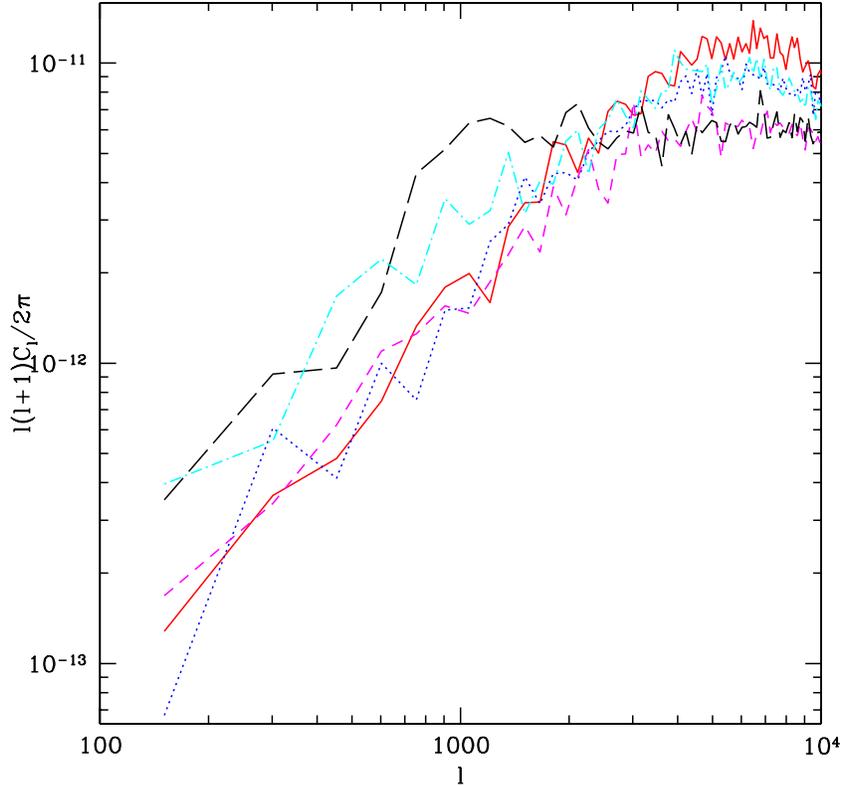,width=4.5in}}
\caption{
Power spectrum variations for several $2^{\circ}\times 2^{\circ}$
fields in $\lambda CDM$ model. There are a factor of 2 variations from 
field to field even on large scales, caused by rare bright objects in 
the map. 
}
\label{fig3}
\end{figure}

Figure \ref{fig3} shows field to field variations in the 
power spectrum 
for $2^{\circ}\times 2^{\circ}$ maps. 
It shows significant fluctuations on large scales.
This is another indication that 
the power spectrum on large scales is dominated by rare bright sources. 
On smaller scales the power spectrum is more robust. The main 
contribution on those scales is from smaller more abundant halos, so 
there is less sampling variance. Note that as the map size increases
the chance of finding a bright source in it increases as well, so the 
median power spectrum on large scales grows with the map size. This is 
equivalent to the power spectrum after the brightest sources have 
been removed. The power spectrum in a 
typical few degree map should correspond to 
the power spectrum with a few thousand brightest sources removed. 
As shown in \cite{KomatsuKitayama99} this can reduce the power spectrum
on large scales significantly and should be kept in mind when 
comparing our results to other predictions. For example,
power spectra in \cite{refregier99}
are based on all the power without the removal of bright sources, which 
should in general give a higher amplitude on large scales. 

\begin{figure}
\centerline{\psfig{file=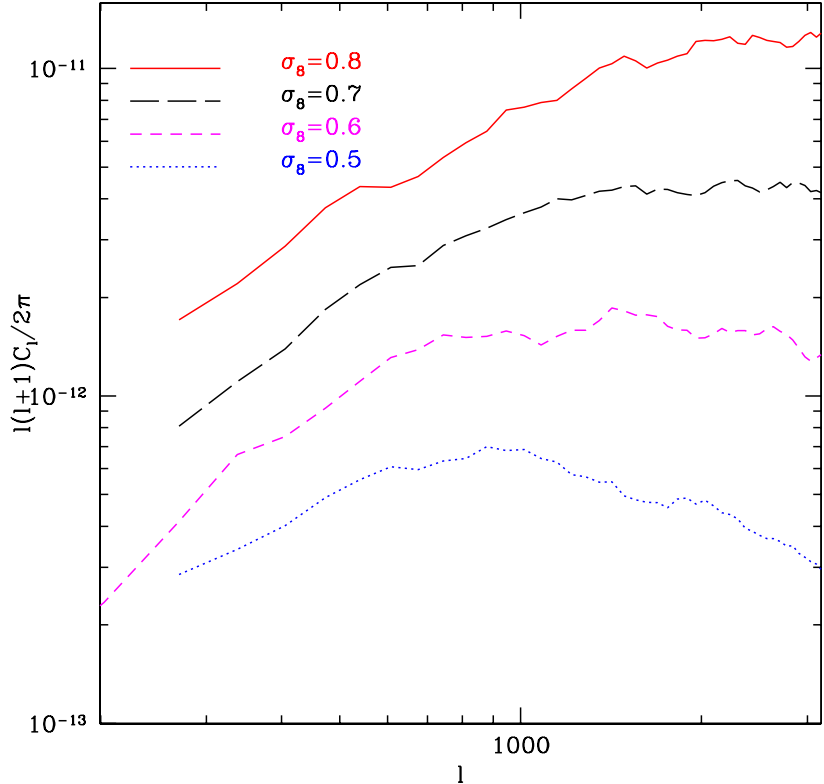,width=4.5in}}
\caption{Dependence of SZ power spectrum on $\sigma_8$ in $\tau CDM$ model. 
}
\label{fig4}
\end{figure}

Figure \ref{fig4} shows the change in the power spectrum as a function 
of $\sigma_8$ for $\tau CDM$ model. The amplitude of the power spectrum
is very sensitive to $\sigma_8$. Doubling this parameter changes the 
power spectrum by 2 orders of magnitude. Fitting to a power law we 
find $C_l \propto \sigma_8^7$. This steep dependence on $\sigma_8$ can 
be understood with Press-Schechter formalism developed in \S 3 and is 
caused by a rapid increase in number of bright sources as a function of
$\sigma_8$, and is in contradiction to the estimates of
\cite{Scaramella93}.  One can also see a crude estimate of the scaling
relation by noting that the $C_l \propto <T> \delta^2$ and the
temperature is a function of the non-linear length scale $r_{\rm NL}$,
i.e. $T
\propto (r_{\rm NL}H_0)^2$ \cite{CO}. For CDM like spectra near the
non-linear mass scale we have $r_{\rm NL}\propto \sigma_8^{1-2}$, giving $C_l
\propto \sigma_8^{6-8}$, similar to what we find in simulations and to 
the PS arguments.
This indicates that care must be exercised when extracting cosmological 
parameters from SZ maps, since similar differences arise between 
low and high density cosmological models (figure \ref{fig2}). A small
increase in $\sigma_8$ of the order of 20\% changes the spectrum as 
much as does changing the density from $\Omega_m=1$ to $\Omega_m=0.3$. 
Unless we are confident that we know the local value of $\sigma_8$ to 
better than this accuracy we cannot use SZ to infer the density of 
the universe. 
Similar argument also explains the redshift dependence of SZ power 
spectrum, which is very strong.
Most of the contribution to the power spectrum comes from 
$z<1$. This is also true in $OCDM$ and $\Lambda CDM$, although the 
$z$ dependence is less steep there.

\subsection{Non-Gaussian signatures}
Since we have 2-d maps of SZ we may use these to study non-Gaussian 
signatures in SZ. 
There are two reasons why to investigate non-Gaussian signatures of SZ.
First is that such information can be used to determine the 
cosmological model. Example of this is the  SZ luminosity 
function, where one identifies SZ sources and plots the number density 
as a function of the flux in SZ \cite{Silva99}. 
The abundance of such sources in the
limit where they are unresolved has been made by the Press-Schechter 
formalism \cite{Aghanim,bar96} or peak-patch formalism \cite{bm96}. 
It has been shown that the number density 
of sources is strongly sensitive to the density parameter, similar to 
the mean $y$ and power spectrum statistics discussed above. 
However, the sources have some internal structure,
they may be clustered, may contain substructure and may not be 
spherical, all of which complicates such analytic approaches and 
they need to be verified using the simulations.
This is of particular interest to the
proposed surveys of small fields centered on a 
random portion of the sky 
\cite{Carlstrom99} which will not in general detect
massive clusters, but rather a collection of smaller sources. 
Simulations such as these presented here
are necessary for investigation of the non-Gaussian effects 
in such random portions of the sky. 

Second reason to study non-Gaussian signatures is that they may 
provide additional leverage in separating the SZ from the primary CMB,
at least under the assumption that primary CMB is Gaussian.
Even when one does not have sufficient frequency 
coverage to distinguish the two components on the basis of their 
different frequency dependence one 
can use non-Gaussian signature of SZ to estimate its contribution 
to the power spectrum. 
The only assumption in this procedure is that at a given smoothing 
scale there is a strong correlation between the non-Gaussian signature
and its second moment, so that the latter can be estimated from the 
former.

\begin{figure}
\centerline{\psfig{file=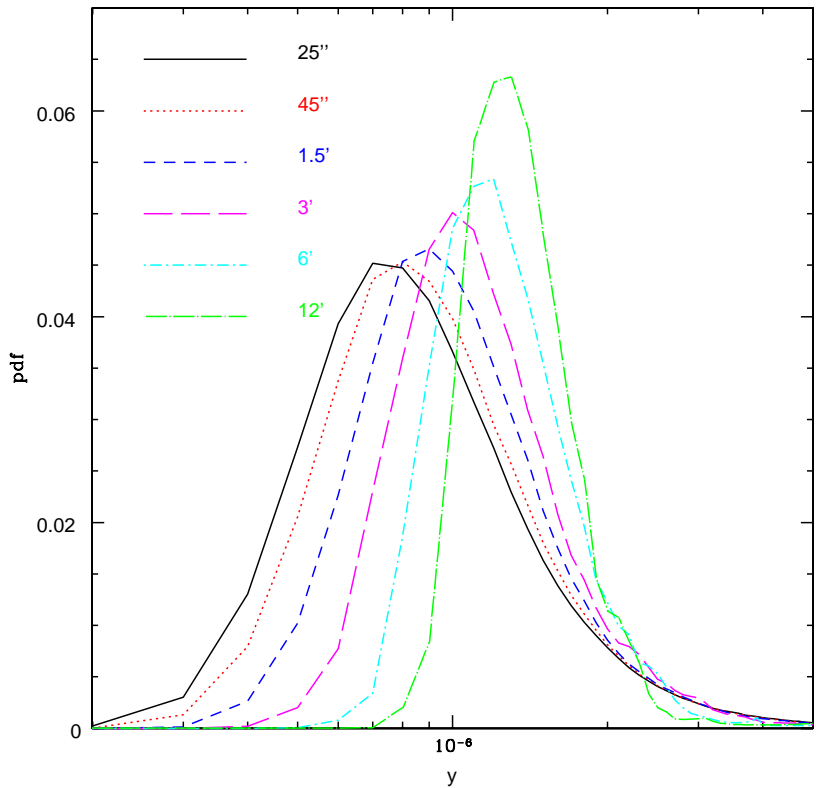,width=4.5in}}
\caption{ 
The one point distribution function of $\log y$ for several 
smoothing radii with a top-hat window.
}
\label{fig5}
\end{figure}

It was shown above that the 2-point statistics is dominated by rare bright 
sources and so is very noisy on large scales. 
This will be even worse if one considers
higher order statistics such as skewness and kurtosis. 
For this reason we concentrate here on the one point distribution function (pdf)
of SZ smoothed at a given angular scale. These are shown in figure \ref{fig5}
for a $2^{\circ}\times 2^{\circ}$ field in $\Lambda CDM$ model with
several smoothing radii, all smoothed with a top-hat window. 
As shown in figure \ref{fig5} the 
pdf is approximately lognormal on small scales, in the sense that there is 
an excess of large $y$ decrements caused by bright rare sources. 
The pdf can be transformed into an
approximate Gaussian if plotted against $\log(y)$.
For larger smoothing 
angles the pdf becomes narrower and approaches a Gaussian, 
so it becomes more difficult to distinguish it 
from primary CMB on this basis.

\subsection{SZ-weak lensing and galaxy cross-correlation}

SZ map can be cross-correlated with other maps, such as weak lensing, 
galaxy or X-ray maps. 
The first two are sensitive to projected density and may not be so 
strongly dominated by rare bright objects, while the latter traces 
projected $\rho^2T^{1/2}$ and is even more
strongly dominated by rare objects than SZ. 
Examples of  
weak lensing maps are those reconstructed from shape distortions 
of $z \sim 1$ galaxies 
or from CMB distortions at $z \sim 1100$. Example of
projected galaxy maps are those from APM, 
SDSS or from numerous degree field surveys. They can be parametrized 
by the mean galaxy redshift.
In the case of SDSS one can use photometric information to weight the 
galaxies according to their distance to optimize the signal to noise of
the cross-correlation. Even more promising are smaller and deeper 
surveys, such as those used for weak lensing studies with several 
hundred thousand galaxies over a degree area. 

\begin{figure}
\vspace*{1cm}
\centerline{\psfig{file=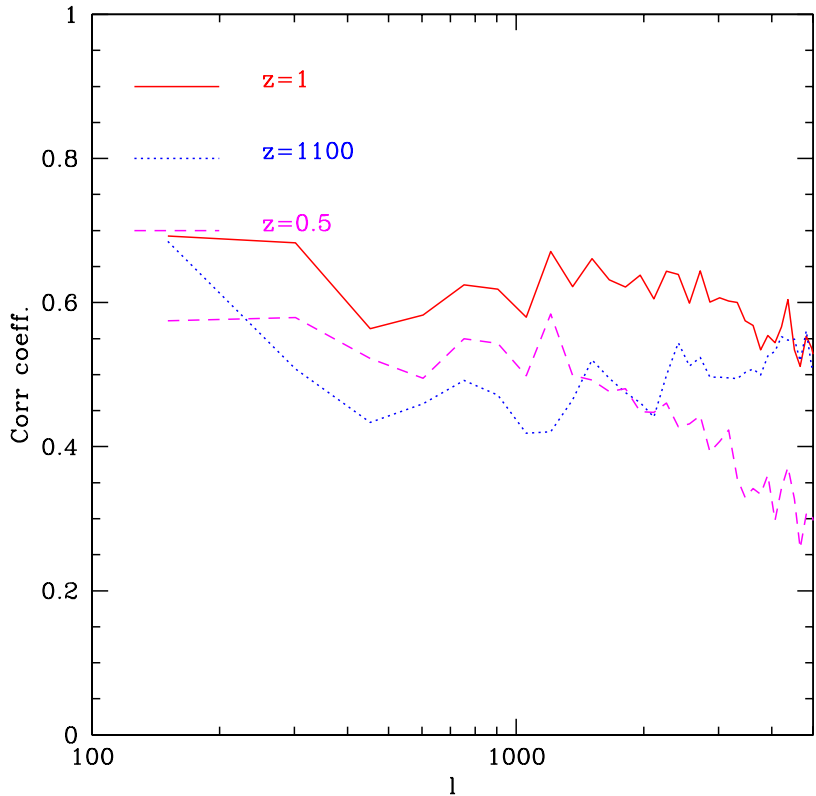,width=4.5in}}
\caption{ 
Cross-correlation coefficient between SZ and weak lensing convergence 
assuming background source is at $z=0.5, 1, 1100$. Similar results are
obtained for cross-correlation with galaxies whose mean distance is at
half the distance to the lens source. 
}
\label{fig6}
\end{figure}

A useful quantity to compare the cross-correlation 
between different maps is to compute the cross-correlation coefficient
\begin{equation}
{\rm Corr}(l)={C_{SZ,X}(l) \over [C_{SZ}(l)C_X(l)]^{1/2}}.
\end{equation}
Figure \ref{fig6} shows the cross-correlation coefficient 
for weak lensing 
map reconstructed from background galaxies at $z=0.5, 1, 1100$. 
The cross-correlation coefficients for all 3 cases is between 
0.4 and 0.6 and only weakly depends on scale. Similar results 
are also obtained if one cross-correlates SZ with a galaxy catalog 
with a mean redshift of $z\sim 0.2-0.5$. This result shows that 
both SZ and weak lensing maps are dominated by relatively nearby 
objects, which is why the cross-correlation coefficient does not 
significantly decrease as the redshift of background galaxies is 
decreased. This is good news for shallow surveys such as SDSS, 
which should be able to detect the cross-correlation signal 
when compared to MAP or Planck maps.

The amplitude of the cross-correlations can also be compared to the analytic
predictions developed in \S3 and we find good agreement between 
the two. The predictions have also been made 
in the case of weak lensing of CMB by
\cite{GoldbergSpergel99,CoorayHu99}.
Our results are several times lower on 
scales between $100<l<3000$.
A more detailed comparison shows that the 
assumed bias in the two models is comparable ($b \sim 3-4$) and
the difference is mostly caused by lower
density weighted temperature in our simulations (0.3keV), consistent with 
PS, compared to their assumption (1keV). Density weighted temperature is 
sensitive to the thermal injection from stars and supernovae and it could 
be significantly higher than in our simulations, which neglect this effect. 
However, such heating of the gas would shift the weight in SZ to
cooler halos (below 1keV), which are not biased or are even antibiased.
So non-thermal increase in density weighted temperature also requires a
decrease in bias and the two effects nearly cancel out, 
leading to a smaller senitivity of the cross-correlation power spectrum
on the density weighted temperature
than one would naively assume. Unfortunately this also implies that the 
expected signal to noise of SZ and CMB cross-correlation 
will be significantly lower 
than predicted \cite{GoldbergSpergel99,CoorayHu99} 
and becomes only marginal using our 
results. 

\section{Discussion and Conclusions}
Using hydrodynamical simulations we have produced maps of SZ and analyzed
some of their low order statistics. The mean Compton parameter was 
found to be in the range $ \bar{y}\sim 1-2.5 \times 10^{-6}$, an order of 
magnitude below current FIRAS limits. Only if one increases $\sigma_8$
by a factor of 2 over the cluster abundance constraint does one violate
the FIRAS limit, something which is clearly excluded based on cluster 
abundance data. 
FIRAS limits will therefore not play a major role in 
constraining the cosmological models. 

The power spectra were found to be noisy and white noise like on large 
scales, indicating that they are dominated by uncorrelated bright 
sources. The amplitude at $l \sim 1000$ is $1-3 \times 10^{-6}$ and does 
not represent a major source of foreground to primary CMB on large 
scales ($l<1000$). On smaller scales SZ power spectrum 
should be detectectable and should dominate
over CMB for $l>2000$. On these scales the predictions become sensitive 
to the thermal energy injection into the gas and  
simulations that ignore such effects 
become unreliable. Conversely, power spectrum of SZ on small scales
should give us important information on the thermal history of the gas
in small halos. Note that the 
dominant contribution to the power spectrum comes
from relatively nearby structures with $z<1$, so SZ will not provide 
information on gas history at high $z$.

On scales where SZ is not negligible compared to primary CMB one
can use the non-Gaussian signatures of SZ to estimate its contribution
to the power spectrum. We have shown that on arcminute scales the 
one point distribution function is well approximated as a 
log-normal, differing significantly from a normal distribution. 
This signature can be used to estimate the contrubution of SZ
to the power spectrum even in the absence of multifrequency 
information.

Further insight into the thermal history of the gas can be obtained
by cross-correlating SZ with other maps that trace large scale
structure, such as weak lensing or projected galaxy map. This can 
also provide information on correlations between groups and clusters on large 
scales and their bias relative to dark matter. The cross-correlation 
coefficient is quite high, of order 0.5, across a wide range of 
scales even when the redshift distribution of galaxies or lensing mass
peaks well below $z \sim 1$. This strong correlation provides further 
incentive to planned SZ surveys on random patches of the sky, 
since even if sensitivity is not sufficient to detect SZ
directly, one may be able to detect it through cross-correlation with weak 
lensing or galaxy maps, thus providing information on the 
thermal state of the gas in the universe.

We thank Eiichiro Komatsu, Alexandre Refregier and David Spergel
for useful correspondence. 
U.S. ackowledges the support of NASA grant NAG5-8084.
Computing support from the National Center for Supercomputing Applications
is ackowledged.
%
%


\end{document}